\documentclass[preprint,superscriptaddress,preprintnumbers,showpacs]{elsart}
\usepackage{amssymb}
\usepackage{dcolumn}
\usepackage{bm}
\usepackage{graphicx}
\usepackage{booktabs}
\usepackage{multirow}

\journal{Physics Letter B. Accepted for publication.}

\begin{document}

\begin{frontmatter}

\title{On self-complementarity relations of neutrino mixing}

\author[1]{Xinyi Zhang},
\author[1,2]{Bo-Qiang Ma\corauthref{*}} \corauth[*]{Corresponding author.}\ead{mabq@pku.edu.cn}
\address[1]{School of Physics and State Key Laboratory of Nuclear Physics and
Technology, Peking University, Beijing 100871, China}
\address[2]{Center for High Energy
Physics, Peking University, Beijing 100871, China}

\begin{abstract}
With the latest results of a large mixing angle $\theta_{13}$ for
neutrinos by the T2K, MINOS and Double Chooz experiments, we find
that the self-complementarity (SC) relations agree with the data in
some angle-phase parametrizations of the lepton mixing matrix. There
are three kinds of self-complementarity relations: (1)
$\vartheta_i+\vartheta_j=\vartheta_k=45^\circ$; (2)
$\vartheta_i+\vartheta_j=\vartheta_k$; (3)
$\vartheta_i+\vartheta_j=45^\circ$ (where $i$, $j$, $k$ denote the
mixing angles in the angle-phase parametrizations). We present a
detailed study on the self-complementarity relations in nine
different angle-phase parametrizations, and also examine the
explicit expressions in reparametrization-invariant form, as well as
their deviations from global fit. These self-complementarity
relations may lead to new perspective on the mixing pattern of
neutrinos.
\end{abstract}


\end{frontmatter}

\newpage
One of the most interesting issues concerning neutrinos is the
misalignment of the flavor eigenstates with the mass eigenstates,
which is the cause of the oscillations as in the neutrino
oscillation theory and is described by the mixing matrix
phenomenologically~\cite{pdg}. Many experiments are set to get
the parameters of the oscillations well determined so as to get
insights into the intriguing nature of
neutrinos~\cite{expt,newexpt,future-expt}.

Just like the Cabibbo-Kobayashi-Maskawa~(CKM) matrix~\cite{CKM}
describing the mixing of quarks, the misalignment of the flavor
eigenstates with the mass eigenstates in the lepton sector can also
be described by a mixing matrix which is called the
Pontecorvo-Maki-Nakagawa-Sakata (PMNS) matrix~\cite{PMNS}. The PMNS
matrix is defined as $U_{\rm PMNS}=U^{l\dagger}_LU^\nu_L$ and can be
expressed generally as
\begin{eqnarray}
 U_{\rm PMNS}=\left(
  \begin{array}{ccc}
    U_{e1}    & U_{e2}    & U_{e3}   \\
    U_{\mu1}  & U_{\mu2}  & U_{\mu3} \\
    U_{\tau1} & U_{\tau2} & U_{\tau3}\\
  \end{array}\right).
\end{eqnarray}

Both the CKM and PMNS matrices can be parameterized by three
rotation angles and a phase angle corresponding to CP violation as
in the standard parametrization i.e. Chau-Keung (CK)
parametrization~\cite{CK}
\begin{eqnarray}
U_{\rm CK}&=&\left(
  \begin{array}{ccc}
    1  & 0     & 0         \\
    0  & c_{23}  & s_{23} \\
    0  & -s_{23} & c_{23} \\
  \end{array}
\right)\left(
  \begin{array}{ccc}
    c_{13}                & 0 & s_{13}e^{-i\delta_{\rm CK}} \\
    0                     & 1 & 0          \\
    -s_{13}e^{i\delta_{\rm CK}} & 0 & c_{13} \\
  \end{array}
\right)\left(
  \begin{array}{ccc}
    c_{12}  & s_{12} & 0 \\
    -s_{12} & c_{12} & 0 \\
    0       & 0      & 1 \\
  \end{array}
\right)\nonumber\\
&=&\left(
\begin{array}{ccc}
c_{12}c_{13} & s_{12}c_{13} & s_{13}e^{-i\delta_{\rm CK}}         \\
-s_{12}c_{23}-c_{12}s_{23}s_{13}e^{i\delta_{\rm CK}} & c_{12}c_{23}-s_{12}s_{23}s_{13}e^{i\delta_{\rm CK}} & s_{23}c_{13} \\
s_{12}s_{23}-c_{12}c_{23}s_{13}e^{i\delta_{\rm CK}} & -c_{12}s_{23}-s_{12}c_{23}s_{13}e^{i\delta_{\rm CK}} & c_{23}c_{13}\\
\end{array}
\right).
\end{eqnarray}
Two additional phase angles are needed for the PMNS matrix if the
neutrinos are of Majorana type. For neutrino mixing, the Majorana
phase angles do not affect the absolute values of the elements of
mixing matrix and are omitted in the above expression.

The angle-phase parametrizations refer to the methods that
parametrize the mixing matrix with three rotation angles and one
(Dirac) or three (Majorana) phase angles. Because a real and
orthogonal matrix can always be decomposed as a product of three
rotations of certain planes, there are options about how to arrange
the orders of these three rotations. Of the twelve ways to do the
product, only nine are independent and the standard parametrization
is one of the nine~\cite{9/12,Zheng10}.

The experimental progresses have upgraded our understanding towards
neutrino mixing in the past several decades. In the attempts of
approximating the PMNS matrix to a constant matrix and proposing the
corresponding theoretical framework on flavor symmetry, the
bimaximal mixing (BM)~\cite{BM}
\begin{eqnarray}
U_{\rm BM}=\left(\begin{array}{ccc}
\sqrt{\frac{1}{2}}    &  \sqrt{\frac{1}{2}}  & 0                \\
-\frac{1}{2}          &  \frac{1}{2}         &\sqrt{\frac{1}{2}}\\
\frac{1}{2}           &  -\frac{1}{2}        &\sqrt{\frac{1}{2}}
\end{array}\right),
\label{BM}
\end{eqnarray}
and tribimaximal mixing (TB)~\cite{TB}
\begin{eqnarray}
U_{\rm TB}=\left(\begin{array}{ccc}
\sqrt{\frac{2}{3}}    &  \sqrt{\frac{1}{3}}  & 0                \\
-\sqrt{\frac{1}{6}}   &  \sqrt{\frac{1}{3}}  & \sqrt{\frac{1}{2}}\\
-\sqrt{\frac{1}{6}}   &  \sqrt{\frac{1}{3}}  &-\sqrt{\frac{1}{2}}
\end{array}\right)
\label{TB}
\end{eqnarray}
have been selected as good description of the data. When working out
the mixing angles in the standard parametrization, both of these two
mixing patterns predict a vanishing mixing angle $\theta_{13}$,
which seems to be challenged by the indication of a relatively large
$\theta_{13}$ in the recent T2K, MINOS and Double Chooz
experiments~\cite{newexpt}. Therefore new perspectives concerning
the mixing pattern of neutrinos are needed to accommodate the new
experimental results.

One interesting direction in the investigation of mixing matrices of
quarks and leptons is trying to find some phenomenological relations
among the mixing parameters, and then trying to find some
theoretical backgrounds or frameworks to understand these relations.
One example is the quark-lepton complementarity~\cite{QLC} between
the mixing angles of quarks and leptons. Different parametrizations
are equivalent to each other mathematically, while some
phenomenological relations are parametrization-dependent. It is
worthwhile to explicitly work out the relations in all the
angle-phase parametrizations as Zheng did in Ref.~\cite{9/12} for
the quark-lepton complementarity, and find some
reparametrization-invariant expressions~\cite{repara}. From the
recent T2K result for a relatively large $\theta_{13}$, there has
been a proposal~\cite{SC} of some self-complementarity relations
among the mixing angles of the neutrino mixing matrix in the
standard parametrization. The purpose of this work is to explore the
phenomenological relations among the mixing angles for the PMNS
mixing matrix, and then check their parametrization-dependence and
-independence in the previously mentioned nine different angle-phase
parametrizations, by confronting with latest experimental results.

We start from a latest global fitting result of the PMNS mixing
matrix~\cite{repara,global},
\begin{eqnarray}
|U_{\rm PMNS}|= \left(
  \begin{array}{ccc}
   0.824^{+0.011(+0.032)}_{-0.010(-0.032)}& 0.547^{+0.016(+0.047)}_{-0.014(-0.044)}
   &  0.145^{+0.022(+0.065)}_{-0.031(-0.113)}    \\
   0.500^{+0.027(+0.076)}_{-0.021(-0.071)}& 0.582^{+0.050(+0.139)}_{-0.023(-0.069)}
   &  0.641^{+0.061(+0.168)}_{-0.023(-0.063)}     \\
   0.267^{+0.044(+0.123)}_{-0.027(-0.088)}& 0.601^{+0.048(+0.133)}_{-0.022(-0.069)}
   &  0.754^{+0.052(+0.143)}_{-0.020(-0.054)}
  \end{array} \right),
\end{eqnarray}
which is obtained from a global fitting of neutrino mixing angles
based on previous experimental data and T2K and MINOS
experiments~($1\sigma~(3\sigma)$)~\cite{global} together with an
Ansatz of a null CP violating phase angle~\cite{repara}. In our
analysis, we calculate the mixing angles of the nine angle-phase
parametrizations with matrix elements that are independent of the
phase angle. For example, from the P1 parametrization, we have,
$$\sin\vartheta_{13}=|U_{e3}|,\quad\tan\vartheta_{12}=\frac{|U_{e2}|}{|U_{e1}|},\quad\tan\vartheta_{23}=\frac{|U_{\mu3}|}{|U_{\tau3}|},$$
thus we get the corresponding values of the mixing angles. The
results are listed in Table~\ref{tab:sc}. Notice that there is no
information on the CP violating phases of both Dirac and Majorana
types for neutrinos from the experiments at present.

\begin{table}[ht]
 \caption{\label{tab:sc}The angle-phase parametrizations and self-complementarity relations}
 \resizebox{\textwidth}{!}{%
 \begin{tabular}{ccc}
   \toprule
 parametrization & $\vartheta_1\slash\vartheta_2\slash\vartheta_3$ & self-complementarity\\
 \hline
 \underline{P1:~$U=R_{23}(\vartheta_{23})R_{13}(\vartheta_{13},\phi)R_{12}(\vartheta_{12})$}& &\\
 \multirow{3}{*}{
  $\left(
   \begin{array}{ccc}
   c_{12}c_{13} & s_{12}c_{13} & s_{13} \\
   -c_{12}s_{23}s_{13}-s_{12}c_{23}e^{-i\phi} &
   -s_{12}s_{23}s_{13}+c_{12}c_{23}e^{-i\phi} &
   s_{23}c_{13} \\
   -c_{12}c_{23}s_{13}+s_{12}s_{23}e^{-i\phi} &
   -s_{12}c_{23}s_{13}-c_{12}s_{23}e^{-i\phi} &
    c_{23}c_{13}\\
   \end{array}
   \right)$}
 &$\vartheta_{12}=(33.58^{+0.577}_{-0.496})^\circ$ &$\vartheta_{12}+\vartheta_{13}=(41.92^{+1.856}_{-2.298})^\circ $\\
 &$\vartheta_{23}=(40.37^{+0.505}_{-0.366})^\circ$
 &\\
 &$\vartheta_{13}=(8.34^{+1.279}_{-1.802})^\circ$
 &$\underline{\vartheta_{12}+\vartheta_{13}\simeq\vartheta_{23}}$\\
 \underline{P2:~$U=R_{12}(\vartheta_3)R_{23}(\vartheta_2,\phi)R_{12}^{-1}(\vartheta_1)$}&
 &\\
 \multirow{3}{*}{
  $\left(
   \begin{array}{ccc}
   s_1 c_2 s_3 +c_1 c_3 e^{-i\phi}& c_1 c_2 s_3-s_1 c_3  e^{-i\phi} & s_2 s_3    \\
   s_1 c_2 c_3-c_1 s_3 e^{-i\phi} & c_1 c_2 c_3+s_1 s_3 e^{-i\phi} & s_2 c_3    \\
   -s_1 s_2                       & -c_1 s_2                     & c_2        \\
   \end{array}
   \right)$}
 &$\vartheta_1=(23.95^{+4.304}_{-2.824})^\circ$
 &$\vartheta_1+\vartheta_3=(36.70^{+6.768}_{-6.716})^\circ$ \\
 &$\vartheta_2=(41.06^{-4.311}_{+1.658})^\circ$
 &\\
 &$\vartheta_3=(12.75^{+2.464}_{-3.892})^\circ$
 &\\
 \underline{P3:~$U=R_{23}(\vartheta_2)R_{12}(\vartheta_1,\phi)R_{23}^{-1}(\vartheta_3)$}&
 &\\
 \multirow{3}{*}{
  $\left(
   \begin{array}{ccc}
   c_1      & s_1 c_3                        & -s_1 s_3   \\
   -s_1 c_2 & c_1 c_2 c_3+s_2 s_3e^{-i\phi}  & -c_1 c_2 s_3+s_2 c_3e^{-i\phi}\\
   s_1 s_2  & -c_1 s_2 c_3+c_2 s_3e^{-i\phi} & c_1 s_2 s_3+c_2 c_3e^{-i\phi} \\
   \end{array}
   \right)$}

 &$\vartheta_1=(34.51^{-0.813}_{+0.739})^\circ$
 &$\vartheta_2+\vartheta_3=(42.95^{+9.597}_{-8.692})^\circ$ \\
 &$\vartheta_2=(28.10^{+5.945}_{-3.441})^\circ$
 &\\
 &$\vartheta_3=(14.85^{+3.652}_{-5.251})^\circ$
 &$\underline{\vartheta_2+\vartheta_3\simeq45^\circ}$\\

 \underline{P4:~$U=R_{23}(\vartheta_2)R_{12}(\vartheta_1,\phi)R_{31}^{-1}(\vartheta_3)$}&
 &\\
 \multirow{3}{*}{
  $\left(
   \begin{array}{ccc}
   c_1 c_3                        & s_1      & -c_1 s_3   \\
   -s_1 c_2 c_3+s_2 s_3 e^{-i\phi}& c_1 c_2  & s_1 c_2 s_3+s_2 c_3 e^{-i\phi}  \\
   s_1 s_2 c_3+c_2 s_3 e^{-i\phi} & -c_1 s_2 & -s_1 s_2 s_3+c_2 c_3 e^{-i\phi} \\
  \end{array}
  \right)$}

 &$\vartheta_1=(33.16^{+1.159}_{-1.014})^\circ$
 &$\vartheta_1+\vartheta_3=(43.14^{+2.914}_{-3.501})^\circ$ \\
 &$\vartheta_2=(45.92^{-0.338}_{+0.160})^\circ$
 &\\
 &$\vartheta_3=(9.98^{+1.755}_{-2.487})^\circ$
 &$\underline{\vartheta_1+\vartheta_3\simeq\vartheta_2\simeq45^\circ}$\\
 \underline{P5:~$U=R_{31}(\vartheta_3)R_{23}(\vartheta_2,\phi)R_{12}^{-1}(\vartheta_1)$}&
 &\\
 \multirow{3}{*}{
  $\left(
   \begin{array}{ccc}
   -s_1 s_2 s_3+c_1 c_3 e^{-i\phi} & -c_1 s_2 s_3-s_1 c_3 e^{-i\phi}  & c_2 s_3   \\
   s_1 c_2                         & c_1 c_2                          & s_2       \\
   -s_1 s_2 c_3-c_1 s_3 e^{-i\phi} & -c_1 s_2 c_3+s_1 s_3 e^{-i\phi}  & c_2 c_3   \\
   \end{array}
   \right)$}

 &$\vartheta_1=(40.67^{+0.796}_{-0.324})^\circ$
 &$\vartheta_2+\vartheta_3=(50.76^{+6.871}_{-4.771})^\circ$ \\
 &$\vartheta_2=(39.87^{+4.943}_{-1.864})^\circ$
 &\\
 &$\vartheta_3=(10.89^{+1.928}_{-2.907})^\circ$
 &\\
 \underline{P6:~$U=R_{12}(\vartheta_1)R_{31}(\vartheta_3,\phi)R_{23}^{-1}(\vartheta_2)$}&
 &\\
 \multirow{3}{*}{
  $\left(
   \begin{array}{ccc}
   c_1 c_3   & c_1 s_2 s_3+s_1 c_2 e^{-i\phi}  & c_1 c_2 s_3-s_1 s_2 e^{-i\phi}   \\
   -s_1 c_3  & -s_1 s_2 s_3+c_1 c_2 e^{-i\phi} & -s_1 c_2 s_3-c_1 s_2 e^{-i\phi}  \\
   -s_3    & s_2 c_3                     & c_2 c_3                      \\
   \end{array}
   \right)$}

 &$\vartheta_1=(31.25^{+1.272}_{-0.960})^\circ$
 &$\vartheta_1+\vartheta_3=(46.74^{+3.921}_{-2.585})^\circ$ \\
 &$\vartheta_2=(38.56^{+0.779}_{-0.484})^\circ$
 &\\
 &$\vartheta_3=(15.49^{+2.649}_{-1.625})^\circ$
 &$\underline{\vartheta_1+\vartheta_3\simeq45^\circ}$\\
 \underline{P7:~$U=R_{31}(\vartheta_3)R_{12}(\vartheta_1,\phi)R_{31}^{-1}(\vartheta_2)$}&
 &\\
 \multirow{3}{*}{
   $\left(
    \begin{array}{ccc}
    c_1 c_3 c_2+s_3 s_2 e^{-i\phi} & s_1 c_3  & -c_1 c_3 s_2+s_3 c_2 e^{-i\phi}  \\
    -s_1 c_2                       & c_1      & s_1 s_2                          \\
    -c_1 s_3 c_2+c_3 s_2 e^{-i\phi}& -s_1 s_3 & c_1 s_3 s_2+c_3 c_2 e^{-i\phi}   \\
    \end{array}
    \right)$}

 &$\vartheta_1=(54.41^{-5.748}_{+2.644})^\circ$
 &\\
 &$\vartheta_2=(52.04^{+1.021}_{+0.707})^\circ$
 &\\
 &$\vartheta_3=(47.69^{+1.885}_{-0.335})^\circ$
 &\\
 \underline{P8:~$U=R_{12}(\vartheta_1)R_{23}(\vartheta_2,\phi)R_{31}(\vartheta_3)$}&
 &\\
 \multirow{3}{*}{
   $\left(
    \begin{array}{ccc}
    -s_1 s_2 s_3+c_1 c_3 e^{-i\phi}& s_1 c_2  & s_1 s_2 c_3+c s_3 e^{-i\phi}   \\
    -c_1 s_2 s_3-s_1 c_3 e^{-i\phi}& c_1 c_2  & c_1 s_2 c_3-s s_3 e^{-i\phi}   \\
    -c_2 s_3                       & -s_2     & c_2 c_3                      \\
    \end{array}
    \right)$}

 &$\vartheta_1=(43.22^{-2.030}_{+0.455})^\circ$
 &$\vartheta_2+\vartheta_3=(56.44^{+6.917}_{-3.800})^\circ$ \\
 &$\vartheta_2=(36.94^{+3.691}_{-1.692})^\circ$
 &\\
 &$\vartheta_3=(19.50^{+3.226}_{-2.108})^\circ$
 &\\
 \underline{P9:~$U=R_{31}(\vartheta_3)R_{12}(\vartheta_1,\phi)R_{23}(\vartheta_2)$}&
 &\\
 \multirow{3}{*}{
   $\left(
    \begin{array}{ccc}
    c_1 c_3  & s_1 c_2 c_3-s_2 s_3 e^{-i\phi}& s_1 s_2 c_3+c_2 s_3 e^{-i\phi} \\
    -s_1     & c_1 c_2                       & c_1 s_2                        \\
    -c_1 s_3 &-s_1 c_2 s_3-s_2 c_3 e^{-i\phi}& -s_1 s_2 s_3+c_2 c_3 e^{-i\phi}\\
    \end{array}
    \right)$}

 &$\vartheta_1=(30.00^{+1.871}_{-1.455})^\circ$
 &$\vartheta_1+\vartheta_3=(47.95^{+5.034}_{-3.371})^\circ$ \\
 &$\vartheta_2=(47.76^{+0.020}_{+0.283})^\circ$
 &\\
 &$\vartheta_3=(17.95^{+3.163}_{-1.916})^\circ$
 &$\underline{\vartheta_1+\vartheta_3\simeq\vartheta_2}$\\
     \bottomrule
\end{tabular}}%
\end{table}

From Table~\ref{tab:sc}, we can see that there are three kinds of
self-complementarity relations (SC)~\cite{SC} satisfied numerically
in five angle-phase parametrizations. Explicit forms of the
self-complementarity relations are underlined in the third column of
the table and can be summarized as follows: \begin{enumerate}
\item
\noindent the first kind, which works in the P4 parametrization,
suggests,
$$\vartheta_i+\vartheta_j=\vartheta_k=45^\circ ; $$
\item
the second kind, which works in the  P1 and P9 parametrizations,
suggests,
$$\vartheta_i+\vartheta_j=\vartheta_k; $$
\item
the third kind, which works in the P3 and P6 parametrizations,
suggests, $$\vartheta_i+\vartheta_j=45^\circ. $$
\end{enumerate}
In the above relations, $i$, $j$ and $k$ denotes different mixing
angles. For the P1 parametrization, which corresponds to the
standard parametrization with a slight difference in phase
convention, we adopt $ij$ to denote different mixing angles as in
most literature. In fact, the first kind relation has been proposed
by Xing in Ref.~\cite{Xing:2010pn} from the viewpoint of adopting a
democratic correction to the tribimaximal pattern, and it has been
also suggested in Ref.~\cite{SC} from pure phenomenological
consideration based on the T2K result.

We start with the first kind of self-complementarity relation in the
P4 parametrization. With
\begin{eqnarray}
R_{\rm 23}(\vartheta_2)=\left(
  \begin{array}{ccc}
    1  & 0          & 0      \\
    0  & c_2     & s_2 \\
    0  & -s_2    & c_2 \\
  \end{array}\right),
  R_{\rm 12}(\vartheta_1,\phi)=\left(
  \begin{array}{ccc}
    c_1  & s_1 & 0 \\
    -s_1 & c_1 & 0 \\
    0       & 0      & e^{-i\phi} \\
  \end{array}\right),
  R_{\rm 31}^{-1}(\vartheta_3)=\left(
  \begin{array}{ccc}
    c_3   & 0 & -s_3 \\
    0     & 1 & 0    \\
    s_3   & 0 & c_3  \\
  \end{array}\right), ~~~~
\end{eqnarray}
we get
\begin{eqnarray}
U&=&R_{23}(\vartheta_2)R_{12}(\vartheta_1,\phi)R_{31}^{-1}(\vartheta_3)\nonumber\\
 &=&\left(
   \begin{array}{ccc}
   c_1 c_3                        & s_1      & -c_1 s_3   \\
   -s_1 c_2 c_3+s_2 s_3 e^{-i\phi}& c_1 c_2  & s_1 c_2 s_3+s_2 c_3 e^{-i\phi}  \\
   s_1 s_2 c_3+c_2 s_3 e^{-i\phi} & -c_1 s_2 & -s_1 s_2 s_3+c_2 c_3 e^{-i\phi} \\
\end{array}
\right).\label{p4}
\end{eqnarray}
Combined with global fitting
results~\cite{global}, we get
\begin{eqnarray}
\vartheta_1=(33.16^{+1.159}_{-1.014})^\circ,\quad
\vartheta_2=(45.92^{+0.160}_{-0.338})^\circ,\quad
\vartheta_3=(9.98^{+1.755}_{-2.487})^\circ.
\end{eqnarray}
Thus we find
\begin{eqnarray}
\vartheta_1+\vartheta_3=(43.14^{+2.914}_{-3.501})^\circ\simeq
\vartheta_2=(45.92^{+0.160}_{-0.338})^\circ\simeq45^\circ.
\end{eqnarray}

As we have seen in Table~\ref{tab:sc}, the self-complementarity
relations are dependent on parametrizations. Though elements of the
mixing matrix cannot be measured independently yet, getting the
expressions of reparametrization-invariant form of the
self-complementarity relations can be used to check the numerical
relations to see how they work in general. This may have the
potential of giving clues to find or build some underlying theories
that can produce such kind of relations.

From Eq.~(\ref{p4}), we have
\begin{eqnarray}
s_1=|U_{e2}|,\quad t_2=\frac{|U_{\tau2}|}{|U_{\mu2}|},\quad
t_3=\frac{|U_{e3}|}{|U_{e1}|}.
\end{eqnarray}
Substituting the trigonometric function with the moduli of the
matrix elements in the following expressions,
\begin{eqnarray}
\tan(\vartheta_1+\vartheta_3)=1,\quad \tan\vartheta_2=1,
\end{eqnarray}
we have,
\begin{eqnarray}
\frac{|U_{e3}|}{|U_{e1}|}+\frac{|U_{e2}|}{\sqrt{1-|U_{e2}|^2}}=
1-\frac{|U_{e3}||U_{e2}|}{|U_{e1}|\sqrt{1-|U_{e2}|^2}};
\end{eqnarray}
\begin{eqnarray}
\frac{|U_{\tau2}|}{|U_{\mu2}|}=1.
\end{eqnarray}
With application of the unitarity relation of the first row of the
PMNS matrix, namely,
$$|U_{e1}|^2+|U_{e2}|^2+|U_{e3}|^2=1,$$
to eliminate $|U_{e2}|$, we can get,
\begin{eqnarray}
\left(\frac{|U_{e1}|-|U_{e3}|}{|U_{e1}|+|U_{e3}|}\right)^2=
\frac{1}{|U_{e1}|^2+|U_{e3}|^2}-1.
\end{eqnarray}

In similar way, we can work out the reparametrization-invariant form
of other kinds of self-complementarity relations as in the P1,
P9, P3 and P6 parametrizations. It is helpful to see how these
expressions work with the global fitting results of the PMNS matrix,
so we list the results in the third column of
Table~\ref{tab:verification}.

\begin{table}[htbp]
 \caption{\label{tab:verification} The reparametrization-invariant forms and their verification with the numerical values from the global fit of the PMNS matrix}
 \begin{tabular}{ccc}
   \toprule
 parametrization & self-complementarity relations  &  numerical values from the global fit \\
  \hline
 P4:&$\frac{1}{|U_{e1}|^2+|U_{e3}|^2}-(\frac{|U_{e1}|-|U_{e3}|}{|U_{e1}|+|U_{e3}|})^2=1$
    &$0.938^{+0.100(+0.296)}_{-0.137(-0.496)}$\\
    &$\frac{|U_{\tau2}|}{|U_{\mu2}|}=1$
    &$1.033^{+0.121(+0.356)}_{-0.056(-0.170)}$ \\
    &\\
 P1:&$\frac{|U_{e3}|(|U_{e1}||U_{\tau3}|+|U_{e2}||U_{\mu3}|)}{\sqrt{1-|U_{e3}|^2}(|U_{e1}||U_{\mu3}|-|U_{e2}||U_{\tau3}|)}=1$
    &$1.231^{+0.651(+1.807)}_{-0.373(-1.221)}$\\
    &\\
 P9:&$\frac{|U_{\mu1}|(|U_{e1}||U_{\mu2}|+|U_{\tau1}||U_{\mu3}|)}{\sqrt{1-|U_{\mu1}|^2}(|U_{e1}||U_{\mu3}|-|U_{\tau1}||U_{\mu2}|)}=1$
    &$1.008^{+0.201(+0.559)}_{-0.109(-0.347)}$\\
    &\\
 P3:&$\frac{|U_{e3}|(|U_{\mu1}|+|U_{\tau1}|)}{|U_{e2}|(|U_{\mu1}|-|U_{\tau1}|)}=1$
    &$0.873^{+0.263(+0.747)}_{-0.236(-0.828)}$\\
    &\\
 P6:&$\frac{|U_{\tau1}|(|U_{e1}|+|U_{\mu1}|)}{\sqrt{1-|U_{\tau1}|^2}(|U_{e1}|-|U_{\mu1}|)}=1$
    &$1.132^{+0.235(+0.657)}_{-0.156(-0.514)}$\\
   \bottomrule
 \end{tabular}
 \end{table}

We see that the self-complementarity relations agree with the data
in the $1\sigma~(3\sigma)$ error range. Besides, we can see that for
central values, the first kind of the self-complementarity relation
in the P4 parametrization agrees well with the data. As it is not an
exact expression, we introduce small deviation angles denoted as
$\alpha$, $\beta$, which satisfy
\begin{eqnarray}
\tan(\vartheta_1+\vartheta_3+\alpha)=1,\quad \tan(\vartheta_2+\beta)=1.\label{deviation}
\end{eqnarray}
From the global fit~\cite{global}, we get
$\alpha=(1.86^{+2.914}_{-3.501})^\circ\quad\beta=(0.92^{+0.160}_{-0.338})^\circ$
in the P4 parametrization and see that $\alpha$, $\beta$ are indeed
small. Notice that Eq.~(\ref{deviation}) is
parametrization-dependent. As the matrix elements are
reparametrization-invariant, it is useful to deduce the expressions
of the deviations and the matrix elements. The following expressions
hold exactly and $\alpha$, $\beta$ can be used to check the
validity and generality of the self-complementarity relations,
\begin{eqnarray}
\frac{1}{|U_{e1}|^2+|U_{e3}|^2}-1=
\left(\frac{(1-\tan\alpha)|U_{e1}|-(1+\tan\alpha)|U_{e3}|}{(1+\tan\alpha)|U_{e1}|+(1-\tan\alpha)|U_{e3}|}\right)^2;
\end{eqnarray}

\begin{eqnarray}
|U_{\tau2}|=\frac{1-\tan\beta}{1+\tan\beta}|U_{\mu2}|.
\end{eqnarray}
For the self-complementarity
relations in other parametrizations, we can introduce similar
parameters to check the deviations of these relations from
experimental global fit.

Several constant matrices are used as the zeroth-order approximation
of the neutrino mixing matrix. Based on the constant matrices,
theories of flavor symmetry are proposed separately. We consider
four constant matrices, namely, bimaximal, hexagonal, tribimaximal
and democratic mixing patterns, all of which have a vanishing
$\theta_{13}$ when working out the corresponding mixing angles in the standard parametrization. Assuming the self-complementarity relation is exact,
we apply it to get corrections to $\theta_{13}$ in the four constant
mixing matrices and list the results in Table~\ref{tab:constant}.

\begin{table}[htbp]
 \caption{\label{tab:constant} The predictions of $\theta_{13}$ from the self-complementarity relation in constant mixing matrices}
 \begin{tabular}{ccc}
   \toprule
 mixing matrix & mixing angles & $\theta_{13}$ from the self-complementarity relation \\
 \hline
 \underline{bimaximal}&&\\
 $\sin^2\theta_{12}=\frac{1}{2}$&$\theta_{12}=45^\circ$&\\
 $\sin^2\theta_{23}=\frac{1}{2}$&$\theta_{23}=45^\circ$&\\
 $\sin^2\theta_{13}=0$          &$\theta_{13}=0^\circ$ &$\theta_{13}=0^\circ$\\
 \hline
 \underline{hexagonal}&&\\
 $\sin^2\theta_{12}=\frac{1}{4}$&$\theta_{12}=30^\circ$&\\
 $\sin^2\theta_{23}=\frac{1}{2}$&$\theta_{23}=45^\circ$&\\
 $\sin^2\theta_{13}=0$          &$\theta_{13}=0^\circ$ &$\theta_{13}=15^\circ$\\
 \hline
 \underline{tribimaximal}&&\\
 $\sin^2\theta_{12}=\frac{1}{3}$&$\theta_{12}=35.264^\circ$&\\
 $\sin^2\theta_{23}=\frac{1}{2}$&$\theta_{23}=45^\circ$&\\
 $\sin^2\theta_{13}=0$          &$\theta_{13}=0^\circ$ &$\theta_{13}=9.736^\circ$\\
 \hline
 \underline{democratic}&&\\
 $\sin^2\theta_{12}=\frac{1}{2}$&$\theta_{12}=45^\circ$&\\
 $\sin^2\theta_{23}=\frac{2}{3}$&$\theta_{23}=54.736^\circ$&\\
 $\sin^2\theta_{13}=0$          &$\theta_{13}=0^\circ$     &$\theta_{13}=9.736^\circ$\\
   \bottomrule
 \end{tabular}
\end{table}

With the T2K results~\cite{newexpt},
$$\theta_{13}^\nu=(9.685^{+4.698}_{-6.289})^\circ~(\mathrm{NH}),\quad
  \theta_{13}^\nu=(10.986^{+5.218}_{-6.848})^\circ~(\mathrm{IH})$$
and global fitting results~\cite{global},
$$\theta_{13}=8.33^\circ\pm1.40^\circ(\pm4.40^\circ),$$
we find that: (1) the self-complementarity correction is compatible
with the data in the error range of T2K results except for the
bimaximal mixing, which satisfies the self-complementarity relation
and needs no correction; (2) the corrections to $\theta_{13}$ in the
tribimaximal mixing and democratic mixing are both close to the
global fit result. There are a number of experiments to measure the
neutrino mixing angle $\theta_{13}$, with primary data
already~\cite{newexpt} or still under data taking
processes~\cite{future-expt}. These future measurements with
improved precision can test the above different predictions.

Also we could start with the self-complementarity relation and seek
for a constant matrix as a new mixing pattern as in Ref.~\cite{SC},
and the resulting matrix can provide better description to the data
than other constant matrices.

In summary, we investigate the self-complementarity relations of
mixing angles of lepton mixing matrix in nine different angle-phase
parametrizations, work out the corresponding reparametrization
invariant expressions, and make some discussions on deviations of
these relations from experimental global fit. We find that the
self-complementarity relations agree with the latest experimental
results and can make compatible prediction when combined with some
available constant matrices. They may also lead to perspective of
new mixing pattern of neutrinos. Better understanding of the
self-complementarity relation may shed light on the mysterious feature
of neutrino mixing and possible underlying theory behind these
phenomenological regularities.

We are very grateful to Ya-juan Zheng and Zhi-zhong Xing for
discussions. This work is partially supported by National Natural
Science Foundation of China (Grants No.~11021092, No.~10975003,
No.~11035003, and No.~11120101004) and by the Research Fund for the
Doctoral Program of Higher Education (China).

{\bf Note added:} There is a novel measurement of the neutrino
mixing angle $\theta_{13}$ by the Daya Bay
Collaboration~\cite{Daya-Bay}, with
$\sin^22\theta_{13}=0.092\pm0.016(\mathrm{stat}) \pm 0.005
(\mathrm{syst})$ of a significance of 5.2~$\sigma$, and the
corresponding angle is
$\theta_{13}=(8.828\pm0.793(\mathrm{stat})\pm0.248(\mathrm{syst}))^\circ$.
The value of $\theta_{13}$ in our analysis is based on the global
fit in Ref.~\cite{global}, with
$\theta_{13}=(8.332\pm1.399(\pm4.396))^\circ$ which
corresponds to $\sin^22\theta_{13}=0.082\pm0.027(\pm0.084)$.
Therefore our analysis are compatible with the new data and we do
not expect deviation from the conclusion in this work.

\end{document}